\documentclass[aps,prl,twocolumn,nofootinbib,twocolumn,floatfix,showpacs,superscriptaddress]{revtex4-2}
\usepackage{amsmath,graphicx,epsfig,amsthm,color,bm}
\usepackage{pifont,bbding,amssymb,soul,mathrsfs,braket}
\usepackage{dcolumn}
\usepackage{siunitx}
\usepackage[hyperindex,breaklinks,colorlinks=true,citecolor=blue,urlcolor=blue]{hyperref}
\usepackage{subfigure}

\usepackage{threeparttable}
\usepackage{booktabs}

\begin{document}

\title{Wavelength-Tunable and High-Heralding-Efficiency Quantum Photon Source in Birefringent Phase-Matched Lithium Niobate Waveguide}
\author{Qi-Tao Zhu}
\email{These authors contributed equally to this work.}
\affiliation{School of Physics, State Key Laboratory of Crystal Materials, Shandong University, Jinan 250100, China}
\author{Xiao-Xu Fang}
\email{These authors contributed equally to this work.}
\affiliation{School of Physics, State Key Laboratory of Crystal Materials, Shandong University, Jinan 250100, China}
\affiliation{Shenzhen Research Institute of Shandong University, Shenzhen} 
\author{He Lu}
\email{luhe@sdu.edu.cn}
\affiliation{School of Physics, State Key Laboratory of Crystal Materials, Shandong University, Jinan 250100, China}
\affiliation{Shenzhen Research Institute of Shandong University, Shenzhen 518057, China}

\begin{abstract}
Lithium niobate~(LN) is a birefringent material, where the strong birefringence thermo-optic effect is promising for the generation of quantum photon source with widely tunable wavelength. Here, we demonstrate birefringent phase-matching in a 20-mm-long waveguide fabricated on 5~$\mu$m-thick x-cut lithium niobate on insulator. The waveguide is deviated from the optical axis of LN by an angle of 53.5$^\circ$, enabling the phase matching between telecom and visible wavelengths. The phase-matching wavelength of this device can be thermally tuned with rate of 0.617~nm/K. We demonstrate the type-1 spontaneous parametric down-conversion to generate photon pairs with brightness of 2.2~MHz/mW and coincidence-to-accidental ratio up to $2.8\times10^5$. Furthermore, the heralded single photon is obtained from the photon pair with efficiency of 13.8\% and count rate up to 37.8~kHz.
\end{abstract}
\maketitle

Quantum photon source is the building block in quantum information processing, where multiple quantum photon sources with identical wavelength is a prerequisite to demonstrate quantum advantages such as quantum computation~\cite{Zhong2020Science}, quantum teleportation~\cite{Yin2012Nature} and quantum internet~\cite{Kimble2008Science,Riedmatten2004PRL,Lu2016Phys.Rev.Lett}. Spontaneous parametric down-conversion~(SPDC)~\cite{Burnham1970Phys.Rev.Lett}, as a second-order nonlinear~($\chi^{(2)}$) optical process, is the most popular technique to generate quantum photon source, including heralded single photon~\cite{Hong1986PRL}, entangled photons~\cite{Kwiat1995PRL} and squeezed light~\cite{Wu1986PRL}.  The two photons generated from SPDC can be engineered to entangled in various degrees of freedom. For instance,  time-bin or time-energy entangled photons can be obtained from SPDC with Franson interference~\cite{JIN2024PQE}, which has been demonstrated on various crystal materials including beta barium borate~(BBO)~\cite{Kwon2013OE}, lithium triborate~(LBO)~\cite{Thew2004PRL}, potassium niobate~(KNbO$_3$)~\cite{Tittel1998PRA}, lithium iodate~(LiIO$_3$)~\cite{Ou1990PRL} periodically poled potassium titanyl phosphate~(PPKTP)~\cite{Shi2006NJP} and periodically poled lithium niobate~(PPLN)~\cite{Martin2017PRL}. However, the phase-matching condition, i.e., energy and momentum conservation, is the fundamental requirement in the design of quantum photon sources, which limits the set of wavelengths of photon sources.

Lithium niobate on insulator~(LNOI) emerges as a promising platform for integrated quantum photon sources~\cite{Elshaari2020NP,Saravi2021AOM,Pelucchi2022NRP,Milad2022AP}. Compared to bulk materials, the ultracompact waveguide on LNOI strongly confines the interacting light that significantly enhances the nonlinear conversion efficiency of SPDC~\cite{Chang2016Optica, Wang2018Optica, Zhao2020Phys.Rev.Lett,Xue2021Phys.Rev.Applied}. On the other hand, LN is a birefringent material, the strong birefringence thermo-optic effect promises to tune the wavelengths of photon sources. Several phase-matching methods have been developed on LNOI. Quasi-phase matching~(QPM) with periodical poling has been widely adopted for efficient SPDC on LNOI~\cite{Zhao2020Phys.Rev.Lett, Javid2021Phys.Rev.Lett, Xue2021Phys.Rev.Applied, Henry2023Opt.Express, Zhang2023Optica,Fang2024Opt.Express}. Very recently, a strict phase-matching technique, so-called modal phase matching~(MPM), has been realized on double-layered LNOI to generate photon pairs~\cite{fang2024_LaserPhotonicsRev._High, Shi2024LightSciAppl}, in which the waveguide is engineered to achieve equal effective refractive indices at desired wavelengths.  

An alternative strict phase-matching technique is known as birefringent phase-matching~(BPM), in which the desired wavelengths with polarizations along different axes are with equal effective refractive indices satisfying type-1 phase matching. The wavelengths satisfying BPM are determined by the angle of wavevectors of light propagating along the waveguide, respect to the optical axis of LNOI. Indeed, BPM on LNOI unitizes the nonlinear susceptibility $d_{31}$ = -4.35 pm/V for SPDC, which is with less efficiency than QPM and MPM utilizing $d_{33}$ = -27 pm/V. Nonetheless, the birefringence thermo-optic effect of LNOI provides the flexibility to thermally adjust the wavelengths of SPDC. The BPM on x-cut LNOI has been realized for second harmonic generation~(SHG)~\cite{Lu2022Opt.Lett,Tang2023Opt.Lett.}, with a wide thermal tunability at a rate of 1.06~nm/K~\cite{Lu2022Opt.Lett}. In contrast to QPM and MPM, BPM does not require preprocessing of LNOI before fabricating the straight waveguide, which is promising to improve the fiber-to-waveguide coupling efficiency using micrometer-thick LNOI~\cite{Zhang2023Optica}.  

In this letter, we report an efficient SPDC source enabled by BPM in a LNOI waveguide. We fabricate a 20-mm-long straight waveguide on a x-cut LNOI with LN thickness of 5~$\mu$m, which deviates the LN optical axis with angle $\theta\approx53.5^\circ$. We characterize the thermal tunability of this device, and demonstrate type-1 SPDC to generate photon pairs and heralded single-photon source~(HSPS).

\begin{figure}[ht]
\centering
\includegraphics[width=\linewidth]{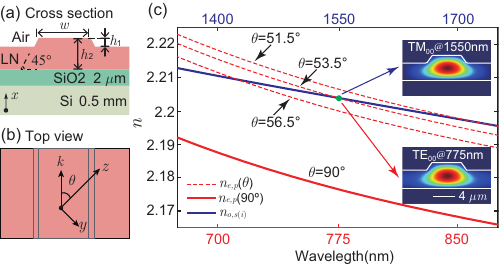}
 \caption{Numerical simulation. (a) The cross-section of the ridge waveguide on $h_2=5~\mu$m thick $x$-cut LNOI wafer, with the top width of $w=5~\mu$m, a sidewall angle of 45$^{\circ}$, an etch depth of $h_1=1.5~\mu$m. (b) The top view of the ridge waveguide and the angle between the waveguide direction~$k$ and the LN optical axis~(z-axis) is $\theta$. (c) The simulated effective refractive indices $n$ provided by the waveguide in (a). The pump light is with fundamental quasi-transverse-electric mode~(TE$_{00}$) and the corresponding $n_{e,p}(\theta)$ depends on $\theta$~(red solid and dashed lines). The signal~(idler) light is with fundamental quasi-transverse-magnetic mode~(TM$_{00}$) and the corresponding $n_{o, s(i)}$ is independent of $\theta$~(blue solid line). The inserts are the mode profiles of pump light at 775~nm and signal~(idler) light at 1550~nm.} 
\label{Fig:design}
\end{figure} 
We consider SPDC taken place in the device of a straight waveguide on LNOI sample consisting 5-$\mu$m-thick $x$-cut LN bonded to 0.5-mm-thick Si substrate with 2-$\mu$m-thick SiO$_2$. As shown in~Figure~\ref{Fig:design}~(a), the waveguide is with top width of $w=5~\mu$m, etch depth of $h_1=1.5~\mu$m and sidewall angle of 45$^{\circ}$. The SPDC is a typical three-wave mixing process, where a high-energy pump photon~($p$) is converted into two lower-energy photons, known as the signal~($s$) and idler~($i$) photons, through $\chi^{(2)}$ nonlinear interaction. In SPDC, phase-matching condition must be satisfied, i.e., $\omega_p=\omega_s+\omega_i$ and $\Delta k = k_s+ k_i - k_p=0$, where $\omega$ is frequency and $k=\frac{2\pi}{\lambda}\cdot n$ is the wavevector and $n$ is the effective refractive index depending on the wavelength $\lambda$. In the waveguide, the pump, signal and idler lights propagate along the $\vec{k}$ direction as shown in~Figure~\ref{Fig:design}~(b), which is deviated by $\theta$ with respect to the optical axis~(z-axis) of LN. In the degenerate case $\lambda_s=\lambda_i=2\lambda_p$, the phase-matching condition $\Delta k = 0$ implies $n_p=n_{s(i)}$. However, the material dispersion would inevitably introduce the phase mismatching. LN is a birefringent material, which is characterized by different refractive indices that depends on the polarization and propagation direction of light. In the case of $\theta=90^\circ$~(the lights propagate along $y$ axis), the lights propagate in fundamental quasi-transverse-magnetic mode~(TM$_{00}$) and quasi-transverse-electric mode~(TE$_{00}$) are ordinary light~(o-light) and extraordinary light~(e-light), respectively. Note that the refractive index of mode TM$_{00}$ is independent of $\theta$, while that of mode TE$_{00}$ depends on $\theta$. This property provides the capability of type-1 SPDC, i.e., $e\to o+o$, by proper design of $\theta$.

To determine $\theta$, we first numerically simulate the effective refractive indices $n_{o,s(i)}$ and $n_{e,p}$ provided by mode TM$_{00}$ and mode TE$_{00}$ at $\lambda_{s(i)}$ and $\lambda_{p}$ in the case of $\theta=90^\circ$. As shown in~Figure~\ref{Fig:design}~(c), the results of $n_{o,s(i)}$ with $\lambda_{s(i)}$ from 1350~nm to 1750~nm are shown with blue solid, and the corresponding $n_{e,p}$ with $\lambda_{p}=\lambda_{s(i)}/2$ are shown with red solid line. It is clearly that $n_{e,p}< n_{o,s(i)}$, so that the phase matching cannot be achieved. In the ideal case, the refractive index $n_{e,p}(\theta)$ is related to $\theta$ by
\begin{equation}\label{Eq:index}
    \frac{1}{n_{e,p}(\theta)^2} = \frac{\sin^2(\theta)}{n_{e,p}(90^\circ)^2} + \frac{\cos^2(\theta)}{n_{o,p}^2}, 
\end{equation}
where $n_{o,p}$ is the effective refractive index provided by TM$_{00}$ at wavelength of $\lambda_p$. According to Equation~\ref{Eq:index}, we calculate $n_{e,p}(\theta)$ with $\theta = 51.5^{\circ}$, 53.5$^{\circ}$ and 56.5$^{\circ}$, and the results are shown with red dashed lines in the Figure~\ref{Fig:design}~(c). The crosses between $n_{e,p}(\theta)$ and $n_{o,s(i)}$ indicate the corresponding $\lambda_{p}$ and $\lambda_{s(i)}$ satisfying the type-1 phase matching. We focus on the SPDC with $\lambda_{p}=$775~nm and $\lambda_{s(i)}=$1550~nm, which corresponds $\theta=53.5^{\circ}$. At $\theta=53.5^{\circ}$, we simulate the mode profiles of TE$_{00}$ at 775~nm and TM$_{00}$ at 1550~nm (inserts in Figure~\ref{Fig:design}~(c)), according to which we obtain $n_{e,p}(53.5^\circ) \approx 2.20405$ and $n_{o,s(i)} \approx 2.20401$.

The fabrication starts with a chip cleaved from an LNOI wafer with 5~$\mu$m thickness of LN~(NANOLN Inc.). We utilize laser-writing to pattern several 20-mm-long waveguides on positive photoresist, which is then transferred to 1-$\mu$m-thick chromium~(Cr) mask using lift-off process and polished by chemical-mechanical polishing~(CMP). The LN is etched using an inductively coupled plasma reactive ion etching tool with argon gas. After etching, a second CMP is performed. Finally, the Cr mask is removed, and the LNOI chip is cleaned by concentrated sulfuric acid. 

\begin{figure}[h!t]
\centering
\includegraphics[width=1\linewidth]{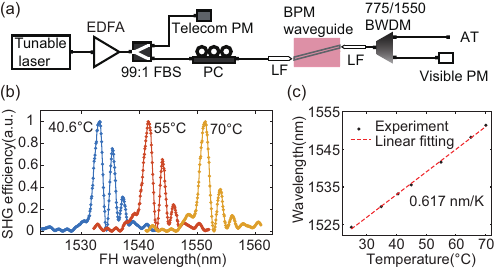}
 \caption{(a) Schematic of the experimental setup for SHG. EDFA, erbium-doped fiber amplifier; 99:1 FBS, 99:1 fibered beamsplitter; PC, polarization controller; LF, lensed fiber; 775/1550 BWDM, 775/1550~nm band wavelength division multiplexer; Visible PM, visible light power meter; AT, absorbing termination. (b) The normalized SHG conversion efficiency spectra at different temperatures. Each spectrum is normalized by its maximal value for clear comparison. (c) The wavelength with maximal SHG efficiency at different temperatures.} 
\label{Fig:shg}
\end{figure} 


\begin{figure*}[h!t]
\centering
\includegraphics[width=1\linewidth]{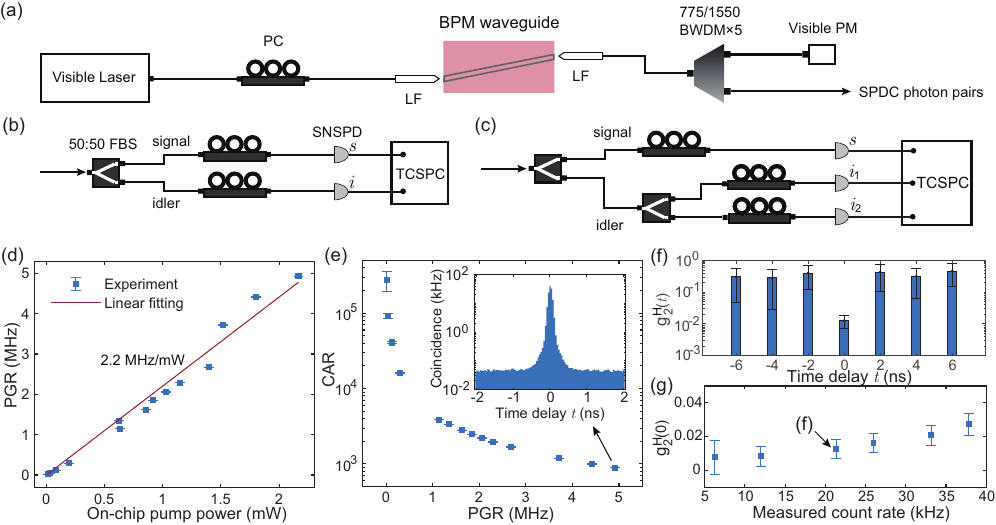}
 \caption{(a) The experimental setups for (a) generation photon pairs via SPDC, (b) characterization of PGR and CAR, (c) characterization the second-order autocorrelation function of HSPS. 50:50 FBS, 50:50 fibered beamsplitter; SNSPD, superconducting nanowire single-photon detectors; TCSPC, time-correlated single-photon counting. (d) PGR at different on-chip pump power. (e) CAR versus PGR. The insert is the coincidence counts between signal and idler channels recorded at different delays $t$ with on-chip pump power of 2.17 mW. (f) $g^{(2)}_\text{H}(t)$ measured at various time delays $t$ with antibunching dip of $g_{H}^{(2)}(0)=0.013\pm0.006$ at on-chip pump power of 0.81~mW. (g) $g^{(2)}_\text{H}(0)$ at different count rates. The error bars are one standard deviation, estimated using Poissonian photon counting statistics.} 
\label{Fig:spdc}
\end{figure*} 

The thermal tunability of device is characterized by SHG with the setup shown in~Figure~\ref{Fig:shg} (a). The first-harmonic~(FH) light in telecom band is provided by a continuous-wave~(CW) tunable laser~(Santec, TSL550) and amplified by an erbium-doped fiber amplifier~(EDFA). Through a 99:1 fibered beam splitter~(FBS), 1\% of the FH light is monitored and the remaining is coupled into the waveguide via a lensed fiber~(LF), where a polarization controller~(PC) is used to ensure the TM polarization of FH light. The LNOI chip is mounted on a thermoelectric cooler~(TEC) for temperature control. The second-harmonic~(SH) light, along with the FH light, is coupled out by another LF, then are separated by a 775/1550~nm band wavelength division multiplexer~(BWDM). A visible power meter~(PM) is used to measure the power of the SH light. By sweeping the wavelength of FH light, we obtain the SHG spectrum as shown in~Figure~\ref{Fig:shg}~(b). We measure the SHG spectrum at different temperatures, and observe the positive temperature dependence of SHG spectrum. Figure~\ref{Fig:shg}~(c) shows the wavelengths with maximal SHG efficiency with respect to the corresponding temperatures, according to which we obtain thermal tunability of the device at rate of 0.617~nm/K.

The experimental setup for SPDC is shown in~Figure~\ref{Fig:spdc}~(a), where the temperature is set at $40.6^{\circ}C$ and the pump light from a CW laser is set at 766.5~nm according to the SHG spectrum shown in~Figure~\ref{Fig:shg}~(b). The pump light is adjusted to TE polarization by a PC for type-1 SPDC, and is coupled into the LNOI waveguide via a LF. The generated signal and idler photons are coupled out from the chip with a second LF, and the pump light is separated by five 775/1550~nm band BWDMs and measured by a visible PM. As shown in~Figure~\ref{Fig:spdc}~(b), we use a 50:50 FBS to separate the signal and idler photons, which are detected by two superconducting nanowire single-photon detectors~(SNSPDs) located in a cryostat~(Quantum Opus, Opus One) and recorded by a time-correlated single-photon counting~(TCSPC) system~(Swabian Instruments, Time Tagger 20) with coincidence window of 1~ns. We denote the count rates of signal photon, idler photon and coincidences as $C_s$, $C_i$ and $C_{si}$, respectively. At each on-chip pump power, we calculate the PGR and CAR respectively. The PGR is calculated by $C_iC_s/2C_{si}$ with the factor of 2 in the denominator being induced by the 50:50 FBS in the experimental configuration for detection. The results of PGR at different on-chip pump power are shown in~Figure~\ref{Fig:spdc}~(d), according to which we linearly fit the data and obtain the brightness of 2.2~MHz/mW. CAR is calculated by $\text{CAR}=\max[g_{si}^{(2)}(t)]-1$ with $g_{si}^{(2)}(t)=C_{si}(t)/C_{si}(\infty)$, where $C_{si}(t)$ is the coincidence count rate at time delay $t$ between signal and idler photons and $C_{si}(\infty)$ is the coincidence at time delay far from $t=0$ as shown in insert of ~Figure~\ref{Fig:spdc}~(e). In our experiment, we observe the CAR ranging from $891 \pm 2.6$ to $278400 \pm 76280$ with corresponding PGR from $4.9$~MHz to $0.33$~kHz. Indeed, the CAR decreases as PGR increases, while their product $\text{CAR}\times \text{PGR}\approx4.4$~GHz is independent of the pump power. 

Triggering one photon, say single photon, heralds the existence of idler photon, which is known as HSPS. The heralding efficiency $\eta^H_i$ of HSPS is determined by~\cite{Klyshk1980,Kaneda2016OE,Henry2023Opt.Express}
\begin{equation}
    \eta_{i}^H = 10^{-\eta_{i}^\text{Loss}/10} = \frac{C_{si}}{C_{s}\cdot \eta_{i}^\text{d}},
\end{equation}
where $\eta_{i}^\text{Loss}$ is the overall loss of idler photon before detection and  $\eta_{i}^\text{d}$ is the detection efficiency of idler photon. In our experiment, the on-chip loss, including the transmission loss and coupling loss, is 3.76~dB. The off-chip loss in fiber devices is 4.82~dB~ (5.09~dB) for signal~(idler) photon. Thus, the overall losses of signal and idler photons are $\eta_{s}^\text{Loss}= $8.58~dB and $\eta_{i}^\text{Loss}= $8.85~dB, corresponding to $\eta^H_s= 13.8$\% and $\eta^H_i=13$\%. We further measure the second-order autocorrelation function of HSPS, known as Hanbury Brown and Twiss effect~\cite{Brown1956Nature}. As shown in Figure~\ref{Fig:spdc}~(c), the idler photon is split by a 50:50 FBS and detected by two detectors $i_1$ and $i_2$. The nonclassical antibunching of HSPS is characterized by second-order autocorrelation function of idler photon at zero time delay~\cite{Zhao2020Phys.Rev.Lett}
\begin{equation}\label{Eq:g2}
    g_{H}^{(2)}(0)=\frac{C_{si_1i_2}C_{s}}{2C_{si_1}C_{si_2}},
\end{equation}
where $C_{si_1i_2}$ is the three-fold coincidence between detectors $s$, $i_1$ and $i_2$. We measure $C_{si_1i_2}$ at different delay time $t$ between $i_1$ and $i_2$, and then calculate $g_{H}^{(2)}(t)$ according to Equation~\ref{Eq:g2}. As shown in~Figure~\ref{Fig:spdc}~(f), a distinct antibunching dip $g_{H}^{(2)}(0)=0.013\pm0.006$ is observed, indicating the small probability of multiphoton emission. Also, $g_{H}^{(2)}(0)$ implies photon-number purity $P$ by $P = 1-g_{H}^{(2)}(0)=0.987\pm0.006$~\cite{faruque2019_Phys.Rev.Applied_Estimating, Shi2024LightSciAppl}.  The count rate of HSPS is proportional to the pump power. However, the high pump power also increases $g_{H}^{(2)}(0)$. In ~Figure~\ref{Fig:spdc}~(g), we show $g_{H}^{(2)}(0)$ and count rate of HSPS at different pump power, which clearly implies the trade-off between the count rate and $g_{H}^{(2)}(0)$. In our experiment, the maximal attainable pump power is 2.34~mW, under which we obtain a HSPS with count rate of 37.8~kHz and $g_{H}^{(2)}(0)=0.027\pm0.006$.

In conclusion, we demonstrate the BPM-enabled type-1 SPDC in a 5~$\mu$m-thick LNOI waveguide. The two-photon generation and HSPS are comprehensively characterized. The thermal tunability of this device is promising for quantum technologies requiring multiply quantum sources with identical wavelength. Indeed, the brightness of two-photon source is 2.2~MHz/mW, which is much lower compared to the devices with LN thickness of hundreds of nanometers~(generally GHz/mW)~\cite{Zhao2020Phys.Rev.Lett, Xue2021Phys.Rev.Applied, Fang2024Opt.Express,fang2024_LaserPhotonicsRev._High,Javid2021Phys.Rev.Lett}, due to the weaker confinement of pump light. Nevertheless, the heralding efficiency is improved to 13.8\%, resulting the measured count rate of HSPS at the same level. The heralding efficiency $\eta^{H}$ is inversely proportional to $g^{(2)}_H(0)$, which is much more important for high-quality HSPS. We expect the heralding efficiency can be further improved by reducing the transition loss, thus benefiting the on-chip HSPS with high brightness and low auto-correlation.  


This work is supported by the National Key R\&D Program of China~(Grants No. 2019YFA0308200), the Shandong Provincial Natural Science Foundation~(Grant No. ZR2023LLZ005), the Taishan Scholar of Shandong Province~(Grants No. tsqn202103013) and the Shenzhen Fundamental Research Program (Grant No. JCYJ20220530141013029)

\bibliography{LNBPM}

\end{document}